\documentclass[twocolumn,prl]{revtex4}
\usepackage{amsfonts}
\usepackage{amsmath}
\usepackage{amssymb}
\usepackage{graphicx}

\begin{document}

\title{Dynamics of particles with "key-lock" interactions}
\author{Nicholas A. Licata and Alexei V. Tkachenko}
\affiliation{Department of Physics and Michigan Center for Theoretical Physics,
University of Michigan, \ 450 Church Str., Ann Arbor, Michigan 48109}

\begin{abstract}
The dynamics of particles interacting by key-lock binding of attached
biomolecules are studied theoretically. \ Examples of such systems include
DNA-functionalized colloids as well as nanoparticles grafted with antibodies
to cell membrane proteins. \ Depending on the coverage of the functional
groups, we predict two distinct regimes separated by a \textit{percolation
transition. \ }In the \textit{localized regime} at low coverage, the system
exhibits a broad, power law like distribution of particle departure times.
At higher coverage, there is an interplay between departure dynamics and
particle diffusion. \ This interplay leads to a sharp increase of the
departure times, a phenomenon qualitatively similar to \textit{aging} in
glassy systems. \ This \textit{diffusive regime} is analogous to dispersive
transport in disordered semiconductors: depending on the interaction
parameters, the diffusion behavior ranges from standard diffusion to
anomalous, \textit{subdiffusive }behavior. \ The connection to recent
experiments and implications for future studies are discussed.
\end{abstract}

\maketitle

Selective key-lock interactions are quintessential for biology. \ Over the
past several years, they have also attracted substantial attention in the
context of nanoscience. \ It is becoming common practice to attach
biomolecules capable of key-lock binding to colloidal particles or other
microscopic objects to achieve controllable, specific interactions. \
Examples include nanoparticles functionalized with complementary
single-stranded DNA (ssDNA) (\cite{crocker}-\cite{colloidalgold}), or with
antibodies to a particular protein. \ The possible applications range from
self-assembly of smart nanomaterials to biosensors and cell-specific drug
delivery (\cite{errorproof}-\cite{membranebend}). \ In this new class of
systems, the collective character of the binding may lead to non-trivial and
often prohibitively slow dynamics. \ 

In this letter we report several remarkable results dealing with the
dynamics of particles with reversible key-lock interactions. \ These results
are of both conceptual and practical interest. \ In particular, we will
demonstrate that depending on the coverage of the functional groups (e.g.
ssDNA or proteins), the system exhibits two distinct regimes separated by a 
\textit{percolation transition}. \ At low coverage, there is a broad power
law like distribution of departure times but no lateral diffusion. If the
coverage is sufficiently high, the overall particle dynamics is a result of
the interplay between diffusion and desorption. The lateral motion is
analogous to \textit{dispersive transport} in disordered semiconductors: it
may range from regular diffusion with a renormalized diffusion coefficient,
to anomalous, subdiffusive behavior. \ 

\begin{figure}[h]
\includegraphics[width=2.7257in,height=2.0475in]{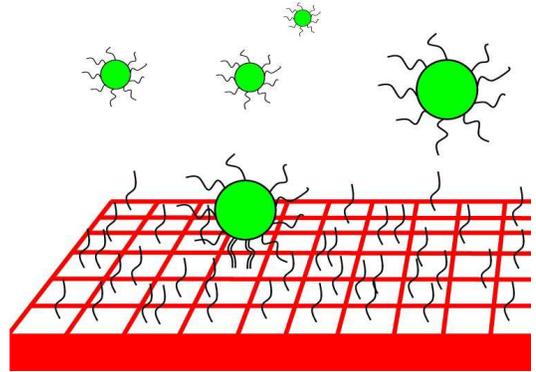}
\caption{(Color online). \ A snapshot of particles interacting with a
two-dimensional substrate. \ Particles are alternately bound to the
substrate by bridges, or unbound and free to diffuse in a direction normal
to the substrate plane. \ }
\label{substrate}
\end{figure}

In the simplest version of our model, a single particle interacts with a
flat $2D$ surface via multiple key-lock binding (see Figure \ref{substrate}%
). \ At each position of the particle, there are $m$ key-lock bridges which
may be closed or open, and there is a binding energy $\epsilon $ for each of
the key-lock pairs (the variation in $\epsilon $ is neglected). Therefore,
the $m$-bridge free energy plays the role of an effective local potential
for the particle: $U\left( m\right) =-k_{B}Tm\Delta $,\ where $\Delta \equiv
\log \left( 1+\exp \left[ \epsilon /k_{B}T\right] \right) $ \cite{statmech}.
\ In a generic case, the number of bridges $m$ is a Poisson distributed
random number $P_{m}=\overline{m}^{m}\exp \left( -\overline{m}\right) /m!$
where $\overline{m}$ denotes the mean of the distribution. \ After staying
for some time at a particular site, the particle either breaks all its
bridges and departs, or hops a distance $a$ to a new site characterized by a
new value for the number of bridges $m$. \ In this sense we have
coarse-grained the particle motion by discrete steps of the correlation
length $a$, the distance after which the number of bridges becomes
statistically independent of the value at the previous location. \ 

There is a percolation transition which separates the diffusive regime from
the localized regime. \ In the localized regime the particle remains close
to the original location until breaking all its bridges and departing. \ In
the diffusive regime the particle undergoes a random walk by breaking and
reforming multiple bridges. \ The transition between the two regimes occurs
at the percolation threshold where one first encounters an infinitely
connected cluster of sites with $m>0$. \ Since the critical probability of
bond percolation on the square lattice\cite{percolation} is $\frac{1}{2}$,
the transition is given by $P_{0}=\frac{1}{2}$. \ \ Below, we calculate the
departure time distribution $\Phi (t)$ in both the localized and diffusive
regimes, and study the random walk statistics above the percolation
threshold. \ 

\textbf{Localized regime}. \ Consider a particle attached to the substrate
by $m$ bridges below the percolation threshold $\overline{m}=\log 2$. \ The
probability that the particle departs from the surface between time $t$ and $%
t+dt$ is $\Phi _{m}(t)dt\simeq K_{m}\exp \left[ -K_{m}t\right] dt$. \ The
departure rate $K_{m}=\frac{1}{\tau _{0}}\exp \left( -\Delta m\right) $ is
given by the Arrhenius relation $K_{m}\sim \exp \left( \frac{U\left(
m\right) }{k_{B}T}\right) $ with $\tau _{0}$ a characteristic timescale for
bridge formation. \ By averaging this distribution over the statistics of $m$
we arrive at:%
\begin{equation}
\Phi (t)=\sum\limits_{m=1}^{\infty }\exp \left[ -K_{m}t\right] K_{m}%
\widetilde{P}_{m}  \label{boundtimenodif}
\end{equation}%
When performing the averaging we do not include states with $m=0$ bridges. \
For this reason we work with a renormalized probability distribution $%
\widetilde{P}_{m}=P_{m}/(1-\exp \left( -\overline{m}\right) )$ so that $%
\sum\nolimits_{m=1}^{\infty }\widetilde{P}_{m}=1$. \ 
\begin{figure}[h]
\includegraphics[width=3.346in,height=2.5192in]{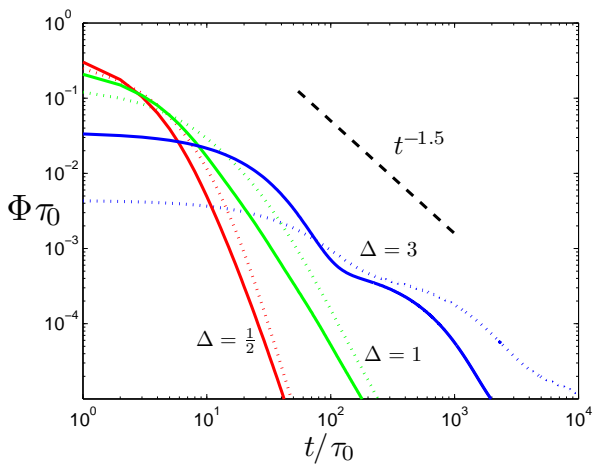}
\caption{(Color online). \ Departure time distribution function vs. time at
the percolation threshold $\overline{m}=\log 2$. \ The solid lines are
calculated from Eq. \protect\ref{boundtimenodif} in the localized regime,
and the dotted curves are calculated from Eq. \protect\ref{phidif} in the
diffusive regime. \ }
\label{percplot}
\end{figure}

The results of this calculation can be compared to a recent experiment which
determined the time-varying separation of two DNA-grafted particles in an
optical trap \cite{crocker}. \ In this experiment two particles are bound by
DNA\ bridges, and after breaking the connections diffuse to the width of the
optical trap. \ Because the length of the DNA\ chains grafted on the
particle is much less than the particle radius, surface curvature effects
can be neglected. \ Hence, the experiment resembles a particle interacting
with a localized site on a $2D$ substrate. \ Experimentally, the tail of the
departure time distribution was observed to be a power law $\Phi (t)\sim
t^{-1.5}$. \ Such behavior is indeed reproduced by Eq. \ref{boundtimenodif}
with a binding free energy on the order of several $k_{B}T$. \ For strong
enough binding, $\Delta \gtrsim 1$, the departure time distribution function
exhibits a pronounced multi-exponential character. \ 

\textbf{Diffusive regime}. \ The departure time distribution is strongly
altered above the percolation threshold. \ In this regime the particle can
move around to find a more favorable state on the surface. This leads to a
much longer lifetime of the bound state, a phenomenon similar to \textit{%
aging} in glassy systems. \ The hopping rate between two neighboring sites
is given by the Arrhenius law, $\kappa _{i\rightarrow j}\sim \frac{1}{\tau
_{0}}\exp \left[ -\Delta (m_{i}-m_{j})\theta \left( m_{i}-m_{j}\right) %
\right] $. \ Here $\theta (x)$ is the Heaviside step function. \ The problem
can be greatly simplified since the ensemble averaged hopping rate from a
site with $m$ bridges can be well approximated by an effective Arrhenius
relation: 
\begin{equation}
\kappa _{m}=\frac{1}{\tau _{0}}\exp \left[ -\Delta (m-\overline{m})\right]
\end{equation}

In the case when $\Delta \overline{m}$ is sufficiently large, the
probability of staying attached to the surface after an $n$ step random walk
is $\left( 1-\frac{K_{m}}{\kappa _{m}}\right) ^{n-1}=\left[ 1-\exp \left(
-\Delta \overline{m}\right) \right] ^{n-1}$. \ Interestingly, this
probability is virtually independent of the particular bridge numbers $%
\{m_{1},...,m_{n}\}$ realized during the walk. \ We \ conclude that the
probability of particle departure after exactly $n$ steps is $f_{n}=\left[
\exp (\gamma )-1\right] \exp (-\gamma n)$, where $\gamma =-\log \left[
1-\exp \left( -\Delta \overline{m}\right) \right] $. \ If we let $\phi
_{n}(t)$ denote the departure time distribution for a walk of $n$ steps we
have: \ 
\begin{eqnarray}
\Phi (t) &=&\sum\limits_{n=1}^{\infty }f_{n}\phi
_{n}(t)=\sum\limits_{n=1}^{\infty
}f_{n}\prod\limits_{i=1}^{n}\sum\limits_{m_{i}=1}^{\infty }\widetilde{P}%
_{m_{i}}\kappa _{m_{i}}\times   \notag \\
&&\int_{0}^{\infty }dt_{i}\exp \left( -\kappa _{m_{i}}t_{i}\right) \delta
\left( t-\sum\limits_{j=1}^{n}t_{j}\right) 
\end{eqnarray}%
To complete the calculation it is convenient to Fourier transform $\phi
_{n}(t)$. \ One can then sum the resulting geometric series to obtain $\Phi
(\omega )$ and perform the inverse transform to derive the following result:%
\begin{align}
\Phi (t)& =\exp (\gamma )\left[ \exp (\gamma )-1\right] \sum\limits_{r=1}^{%
\infty }\frac{\exp (-z_{r}t)}{Y(z_{r})}  \label{phidif} \\
Y(z_{r})& =\sum\limits_{m=1}^{\infty }\frac{\widetilde{P}_{m}\kappa _{m}}{%
\left( \kappa _{m}-z_{r}\right) ^{2}}
\end{align}%
Here $z_{r}$ labels the roots of the equation%
\begin{equation}
\exp (\gamma )-\sum\limits_{m=1}^{\infty }\frac{\widetilde{P}_{m}\kappa _{m}%
}{\kappa _{m}-z}=0.
\end{equation}

In Figure \ref{percplot} the two results for the departure time distribution
are compared at the percolation threshold $\overline{m}=\log 2$. \ For fixed 
$\overline{m}$ a change in $\Delta $ is directly related to a change in the
average binding free energy. \ As indicated in the plot, increasing $\Delta $
decreases the rate of particle departure. \ 

\begin{figure}[h]
\includegraphics[width=3.346in,height=2.5192in]{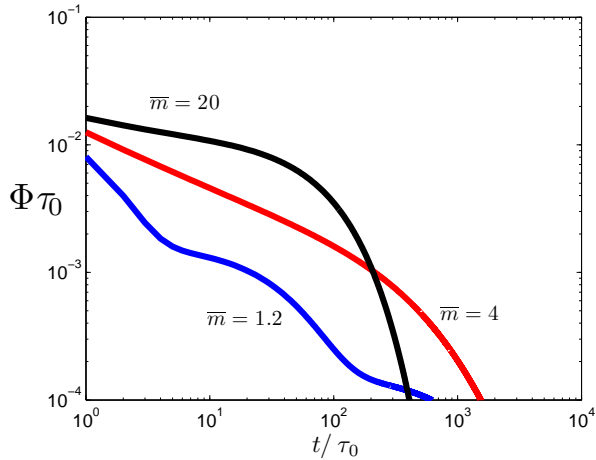}
\caption{(Color online). $\ $Departure time distribution function vs. time
as determined by Eq. \protect\ref{phidif} in the diffusive regime. $\ $In
the plot $\Delta \overline{m}=4$. \ }
\label{difplot}
\end{figure}

In Figure \ref{difplot} we plot the departure time distribution as
determined by Eq. \ref{phidif} in the diffusive regime. \ In the figure we
hold the product $\Delta \overline{m}=4$ constant. \ The optimal regime for
fast departure is to have a large number of weakly bound bridges. \ In this
case the departure time is well approximated as a single exponential, $\Phi
(t)=K_{\overline{m}}\exp (-K_{\overline{m}}t)$. \ 

Finally, we discuss the statistics of the in-plane diffusion of the
particle. We notice that the in-plane \textit{trajectory }of the particle
subjected to a delta-correlated random potential remains statistically
equivalent to an unbiased random work. Therefore, the mean-squared
displacement after $n$ steps is given by $\left\langle r^{2}\right\rangle
=na^{2}$. However, as the particle explores the landscape the average
hopping time becomes longer and the diffusion gets slower. \ In the limit $%
n\rightarrow \infty $, the average hopping time can be determined from the
equilibrium canonical distribution. \ For the case of Poisson distributed $m$%
, this corresponds to a finite yet renormalized diffusion coefficient $%
D^{\ast }$ with $D_{0}=a^{2}/4\tau _{0}$: 
\begin{equation}
D^{\ast }\equiv \frac{1}{4}\frac{\partial \left\langle r^{2}\right\rangle }{%
\partial \left\langle t\right\rangle }=D_{0}\frac{\exp \left( \overline{m}%
e^{\Delta }\right) -1}{\exp \left( \Delta \overline{m}\right) \left[ \exp
\left( \overline{m}\right) -1\right] }
\end{equation}%
\ 

\begin{figure}[tp]
\includegraphics[width=3.346in,height=2.5192in]{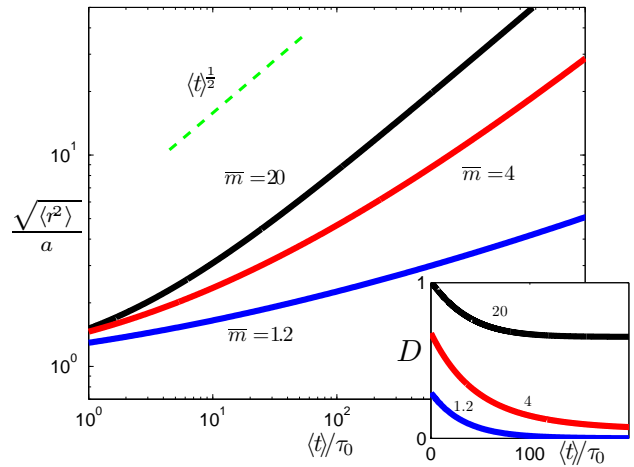}
\caption{(Color online). \ Rms displacement vs. time holding $\Delta 
\overline{m}=4$. \ The inset is the dimensionless diffusion coefficient
defined as $D=\frac{1}{4D_{0}}\frac{\partial \left\langle r^{2}\right\rangle 
}{\partial \left\langle t\right\rangle }$ plotted against time. \ }
\label{r2t}
\end{figure}

However, it may take a very long time to achieve this "ergodic" behavior. \
In the transient regime, an $n-$step random walk cannot typically visit
sites with an arbitrarily large number of bridges $m$. \ Instead, one should
average the hopping times only over sites with $m<m^{\ast }$. \ In the
language of the statistics of extreme events, $m^{\ast }-1$ is the maximum
"expected" value of $m$ in a sample of $n$ independent events \cite%
{dispersive}. Both the average diffusion time $\left\langle t\right\rangle $%
, and mean square displacement $\left\langle r^{2}\right\rangle =na^{2}$,
can be expressed in terms of $m^{\ast }$, which defines their relationship
in parametric form:%
\begin{equation}
\left\langle r^{2}\right\rangle =\frac{a^{2}}{P(\overline{m},m^{\ast })}
\label{mstar}
\end{equation}%
\ 
\begin{equation}
\left\langle t\right\rangle =\frac{\left\langle r^{2}\right\rangle }{D^{\ast
}}\left( 1-\frac{P(\overline{m}e^{\Delta },m^{\ast })}{1-\exp \left( -%
\overline{m}e^{\Delta }\right) }\right)  \label{timenonerg}
\end{equation}%
Here $P(x,m^{\ast })\equiv \gamma (x,m^{\ast })/\Gamma \left( m^{\ast
}\right) =\exp \left( -x\right) \sum\limits_{k=m^{\ast }}^{\infty }x^{k}/k!$
is the regularized lower incomplete $\Gamma $ function. It is easy to see
that in the limit $m^{\ast }\rightarrow \infty $ we recover the renormalized
diffusion relation $\left\langle t\right\rangle =\left\langle
r^{2}\right\rangle /D^{\ast }$, although this occurs at very long, often
unrealistic times. \ In the transient regime we expect anomalous,
subdiffusive behavior. As shown in Figure \ref{r2t}, this regime is typical
for strong enough key-lock interactions. The predicted anomalous diffusion
may be well described by a power law, $\left\langle r^{2}(t)\right\rangle
\sim \left\langle t\right\rangle ^{\alpha }$, with a non-universal exponent $%
\alpha <1$. \ 

\begin{figure}[tph]
\includegraphics[width=3.011in,height=2.267in]{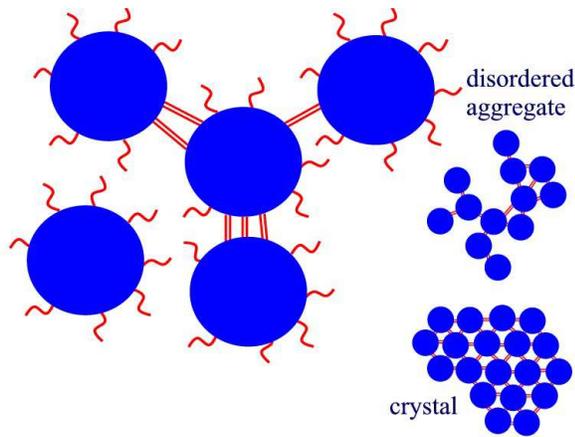}
\caption{(Color online). \ Schematic depiction of key-lock binding between
nanoparticles functionalized with complementary ssDNA. \ The resulting
structures can be disordered, fractal-like aggregates, or crystalline. \ }
\label{crystal}
\end{figure}

\begin{figure}[tph]
\includegraphics[width=3.011in,height=2.267in]{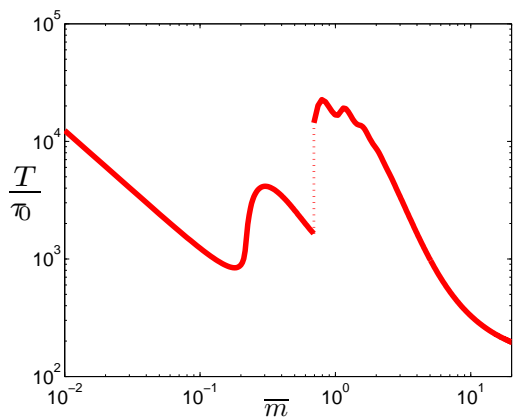}
\caption{(Color online). \ Plot of the time $T$ required for $90\%$ of the
particles to depart vs. $\overline{m}$. \ Note the jump in $T$ which occurs
at the percolation threshold separating the localized regime from the
diffusive regime. \ }
\label{T90}
\end{figure}

This work provides additional insight into the slow crystallization dynamics
of key-locking particles (see figure \ref{crystal}). \ In \cite{chaikin}, $1$
$\mu m$ diameter particles were grafted with ssDNA and formed reversible,
disordered aggregates. \ The average number of key-lock bridges between
particle pairs was $\overline{m}\sim 1$. \ By further reducing the grafting
density of DNA strands on the particles ($\overline{m}<1$), the authors of
reference \cite{crocker} observed random hexagonal close-packed crystals. \
Crystallization requires that colloids repeatedly depart and reattach to the
growing structure, in an effort to find their desired lattice location. \ We
can attempt to quantify this optimal experimental regime of fast departure
by determining the time $T$ required for 90\% of the particles to depart: $%
0.1=\int_{T}^{\infty }\Phi (t)dt$. \ Figure \ref{T90} is a plot of $T$ vs. $%
\overline{m}$ at constant binding free energy \cite{statmech}: $\frac{\Delta 
\overline{m}}{1-\exp (-\overline{m})}+\log (1-\exp (-\overline{m}))=const$.
\ The optimal regime is to have a large number ($\overline{m}\sim 10$) of
weakly bound bridges. \ To realize this regime experimentally we propose the
introduction of long, flexible DNA linkers to a system of particles with a
high coverage of short ssDNA. \ This scheme increases the number of key-lock
bridges between particle pairs as compared to previous experiments, and
therefore has the potential to substantially reduce the time required for
crystallization. \ The prediction of a localized regime below the
percolation threshold (the local minima in Figure \ref{T90}) where particle
departure is relatively fast is confirmed experimentally by the
crystallization observed in \cite{crocker}. \ \ \ 

In this work we studied the dynamics of particles which form multiple,
reversible key-lock bridges. \ There is a \textit{percolation transition}
which separates the regime in which particles are localized near their
original location from the regime where they exhibit diffusive behavior by
breaking and reforming bridges. \ At low coverage the key-locking system
exhibits a broad, power law like distribution of departure times, but no
lateral diffusion. \ Above the percolation transition ($\overline{m}=\log 2$%
) diffusion allows the particle to cascade into deeper energy wells with a
large number of key-lock bridges. \ This leads to an increase in the bound
state lifetime similar to \textit{aging} in glassy systems. \ The statistics
for the particles' in-plane diffusion were determined. \ For relatively weak
key-lock interactions there is a finite renormalization of the diffusion
coefficient. \ However, as $\Delta $ increases, the system exhibits
anomalous, sub-diffusive behavior analogous to dispersive transport in
disordered semiconductors. \ We discussed the connection between our work
and recent experiments with DNA-coated colloids. \ The findings indicate
that the optimal regime for colloidal crystallization, where particle
departure is a relatively fast process, is to have a large number of weakly
bound key-lock bridges. \ 

This work was supported by the ACS Petroleum Research Fund (PRF Grant No.
44181-AC10). \ We acknowledge L. Sander, B. Orr, and B. Shklovskii for
valuable discussions. \ 

\bibliographystyle{achemso}
\bibliography{acompat,dna}

\end{document}